\documentclass[prd,showpacs,showkeys,floatfix,twocolumn,amsmath]{revtex4}
\usepackage{graphicx,dcolumn,booktabs,bm}
\usepackage{longtable,lscape}
\usepackage{amssymb}
\usepackage{indentfirst}
\usepackage{epsfig}
\usepackage{feynmf}   
\usepackage{epstopdf}   
\usepackage{slashed}  
\usepackage{cases}
\usepackage{multirow}
\newcommand {\tabincell}[2]{\begin{tabular}{@{}#1@{}}#2\end{tabular}}%
\newcommand{\BR}{{\cal B}}
\newcommand{\EE}{e^+e^-}

\begin{document}

\title{Sensitivity Study of Searching for $\tau^- \to \gamma \mu^-$ at HIEPA}

\author{Yu-Bo Li$^{1}$}
\author{Cheng-Ping Shen$^{1}$}
\email{shencp@buaa.edu.cn}
\author{Chang-Zheng Yuan$^{2}$}
\affiliation{
$^1$School of Physics and Nuclear Energy Engineering,
Beihang University, Beijing 100191, China \\
$^2$Institute of High Energy Physics, Chinese Academic of Science,
Beijing 100049, China}

\begin{abstract}

The charged lepton flavor violation process is a clean and sensitive
probe of new physics beyond the Standard Model. A sensitivity
study is performed to the process $\tau^- \to \gamma \mu^-$ based
on a 3~fb$^{-1}$ inclusive Monte Carlo sample of $e^+e^-$
collisions at a center-of-mass energy of 4.26 or 4.6~GeV, in the
framework of the BESIII software system. The 90\% confidence level
upper limits on $\BR(\tau^- \to \gamma \mu^-)$ are estimated
assuming no signal is produced. We also obtain the sensitivity on
$\BR(\tau^- \to \gamma \mu^-)$ as a function of the integrated luminosity,
to serve as a reference for the HIEPA being proposed in China.
It is found that 6.34~ab$^{-1}$ are needed to reach the
current best upper limit of $4.4\times 10^{-8}$ and about
2510~ab$^{-1}$ are needed to reach a sensitivity of $10^{-9}$
if the detector design is similar to that of BESIII.

\end{abstract}



\maketitle

\section{Introduction}
\label{sec:intro}

Lepton flavor violation (LFV) in charged lepton decays is
forbidden in the Standard Model (SM) and is highly suppressed even
if mixing between neutrino flavors is taken into
account~\cite{SM} since the rates of LFV processes are suppressed
by the fourth power of $\frac{m_{\nu}}{m_{W}}$, where $m_{\nu}$
and $m_{W}$ are the masses of neutrino and $W$ boson~\cite{PDG},
respectively. On the other hand, the rates of LFV may be enhanced
to observable level in various new physics scenarios beyond the
SM, such as the Minimal Supersymmetric extension of the SM
(MSSM)~\cite{SSM}, Grand Unified Theories~\cite{grand}, and seesaw
mechanisms~\cite{SEESAW}.

In these models, $\tau$ decay is an ideal probe to new physics
because it is the heaviest charged lepton with many possible LFV
decay modes. The branching fractions from the model predictions
are in a range of $10^{-9} \sim 10^{-7}$~\cite{review}, which are
as high as the experimental sensitivity in current B-factory
experiments, and the radiative decays $\tau^-\to \gamma\mu^-$ and
$\tau^-\to \gamma e^-$ are regarded as golden
channels~\cite{footnote}.
LFV process can also be
searched for in $\mu^- \to e^-$ conversion.
SINDRUM II Collaboration studied $\mu^- \to e^-$ conversion in a muonic atom,
giving $R_{\mu e} = \sigma(\mu^- \hbox{Au} \to e^- \hbox{Au})/ \sigma(\mu^- \hbox{Au} \to \hbox{capture})~<$ 10$^{-13}$
at a 90\% confidence level (C.L.)~\cite{mue1}. Future Mu2e experiment is expected to reduce the upper limit of
$R_{\mu e}$ to 6 $\times$ 10$^{-17}$~\cite{mue2}.

Observation of LFV will be a clear signal of new physics, it
directly addresses the physics of flavor and of generations. The
searches for LFV have been a long history (for a review, see
Ref.~\cite{review}), however, no evidence has ever been observed. The best upper limits at a 90\% confidence level are $\BR(\tau^-\to
\gamma \mu^-)<4.4\times 10^{-8}$ and $\BR(\tau^-\to \gamma
e^-)<3.3\times 10^{-8}$, obtained by the BaBar experiment using 963
million $\tau$ decays~\cite{babar}.

\begin{figure}[htbp]
  \includegraphics[width=0.23\textwidth]{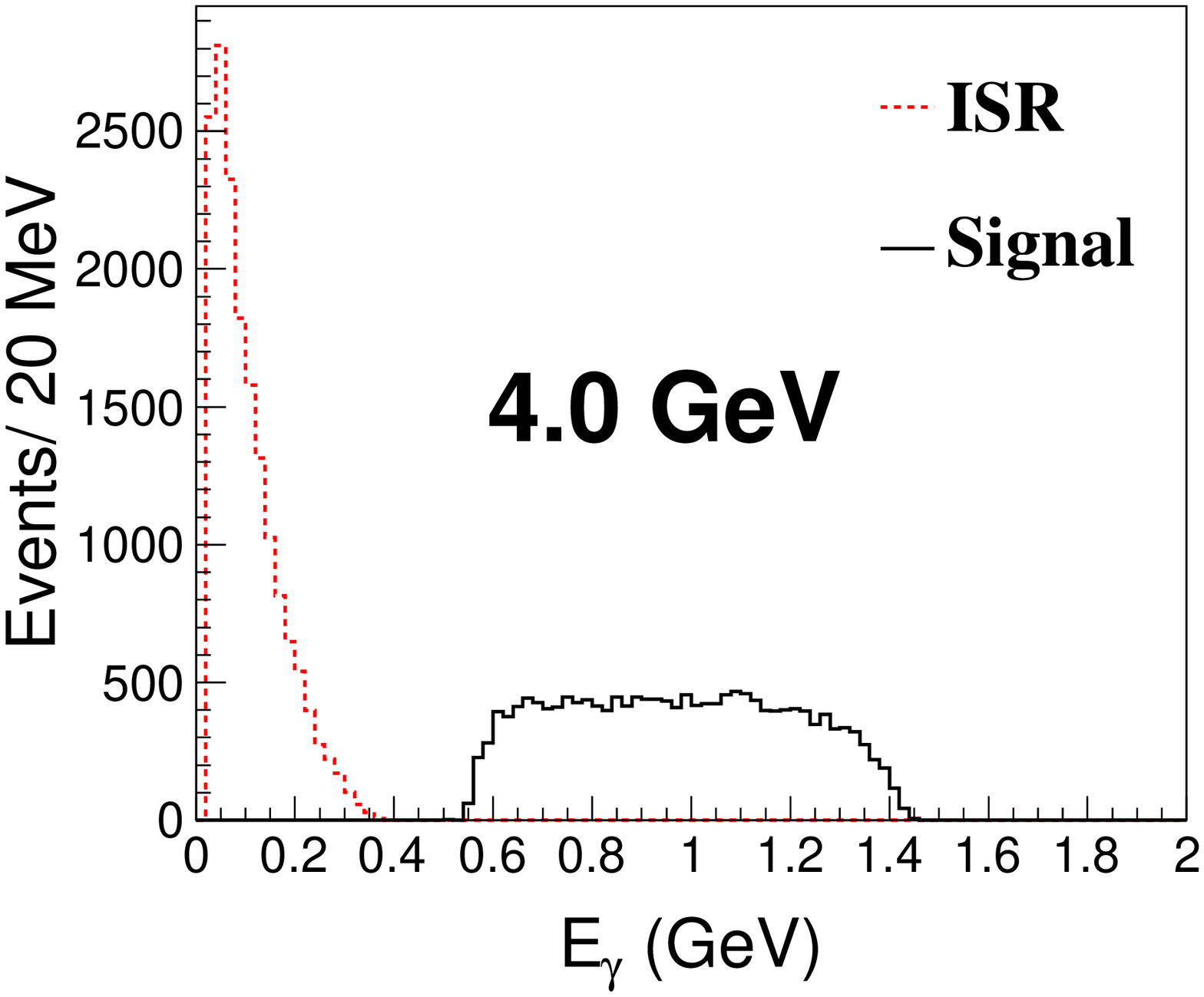}
  \includegraphics[width=0.23\textwidth]{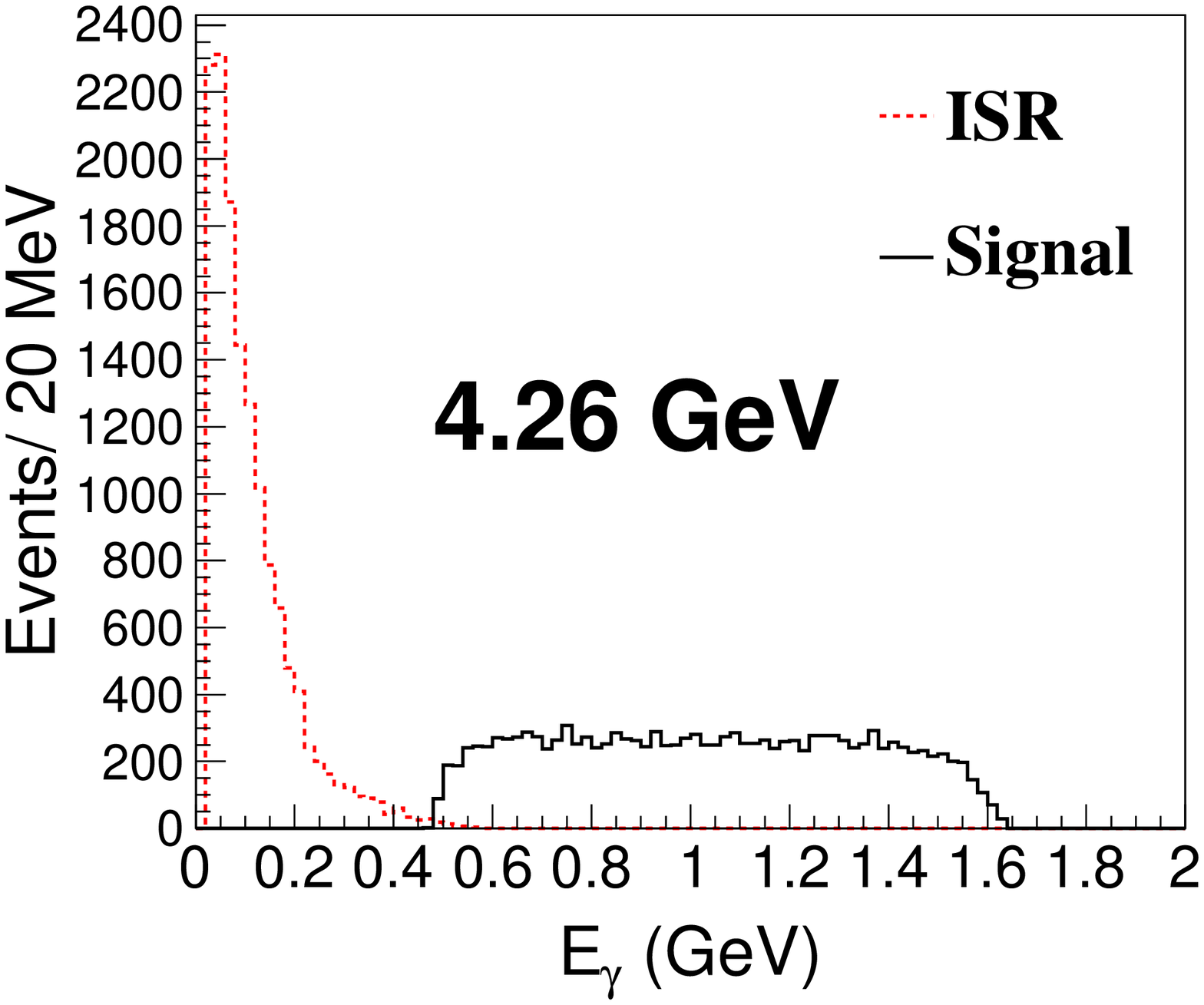}\\
  \includegraphics[width=0.23\textwidth]{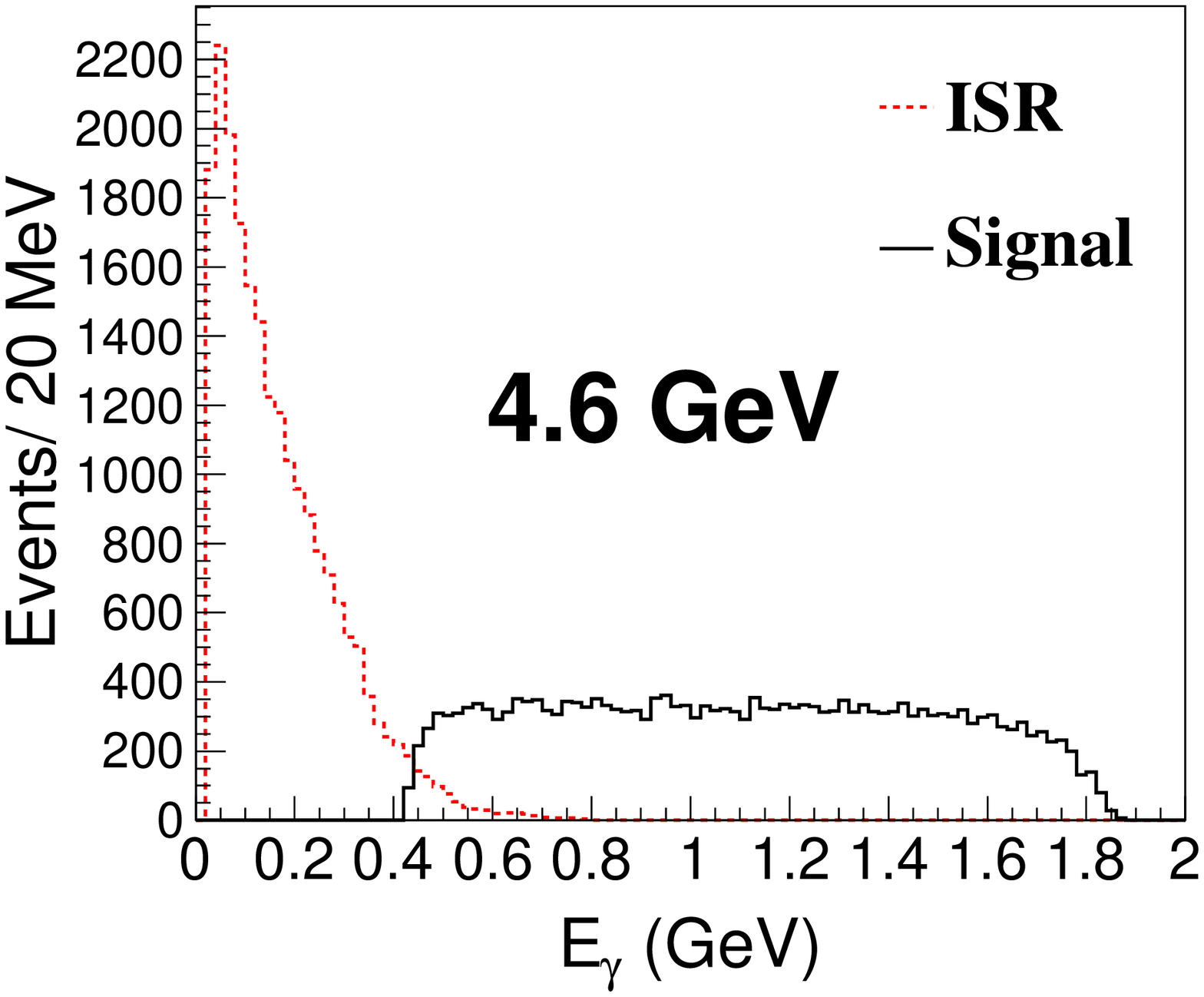}
  \includegraphics[width=0.23\textwidth]{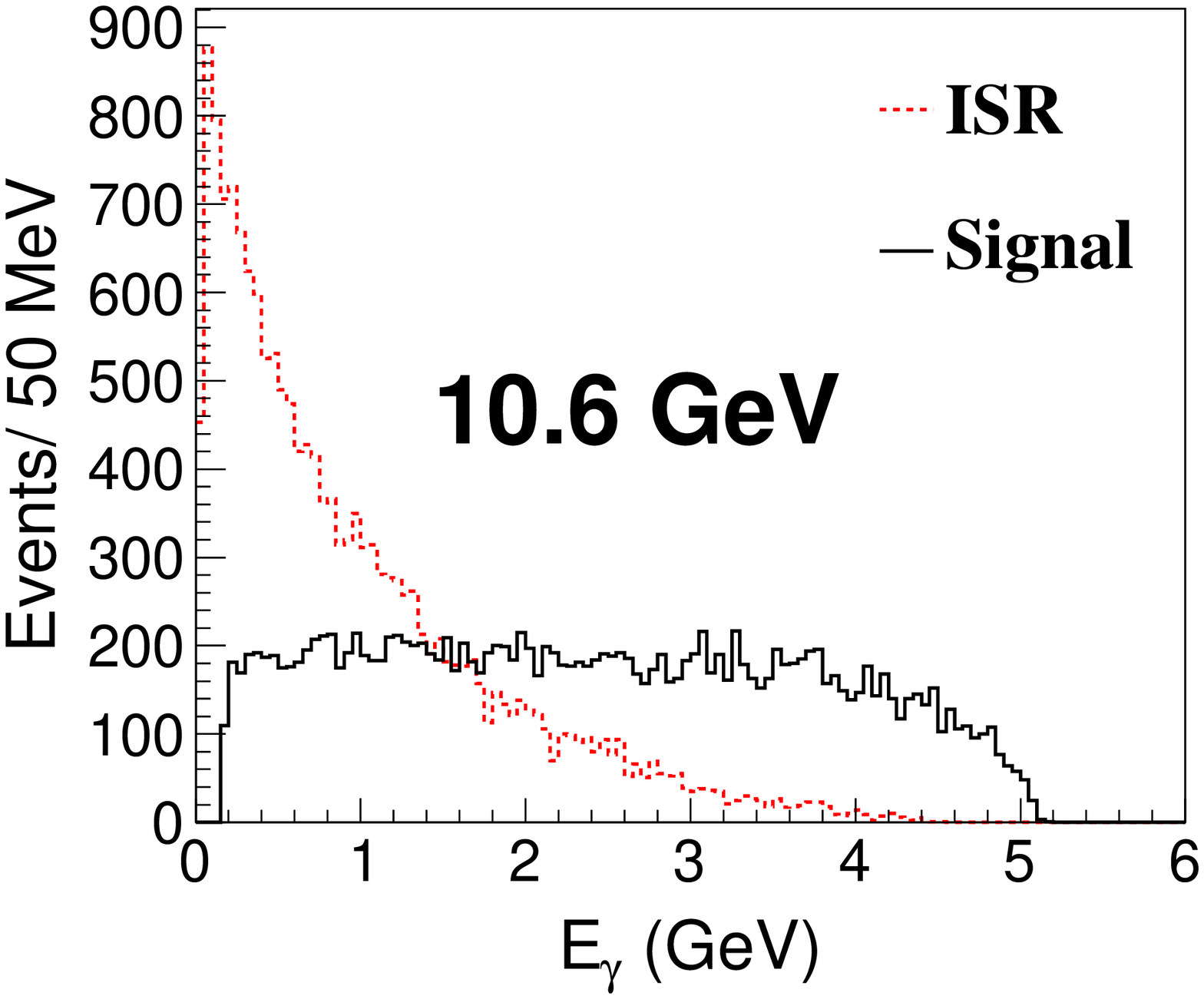}
\caption{The photon energy distributions in $e^+e^-$ CM system
from MC simulated samples $e^{+} e^{-} \to \gamma_{\rm ISR}
\tau^{+} \tau^{-}$, $\tau^{-} \to \gamma_{\rm signal} \mu^{-}$,
$\tau^{+} \to anything$ at $\sqrt{s}=4.0$ (top left), 4.26 (top right), 4.6 (bottom
left), and 10.6~GeV (bottom right). The dashed
histograms are photons from ISR ($\gamma_{\rm ISR}$), while the
solid ones represent the radiative photon from $\tau$ decays
($\gamma_{\rm signal}$).} \label{ISR}
\end{figure}

For a $\tau^-\to \gamma \mu^-$ search at the B-factories, the
dominant background originates from $\EE\to \tau^+\tau^-$ with
initial state radiation (ISR), i.e., $\EE\to \gamma_{\rm
ISR}\tau^+\tau^-$, where one of the $\tau$s decays
semi-leptonically and the final state lepton and the ISR photon
compose signal candidates~\cite{babar}. Such a background
can be avoided at the lower center-of-mass (CM) energy ($\sqrt{s}$) at
a $\tau$-charm factory. Figure~\ref{ISR} shows the photon energy
distributions for $\sqrt{s}=4.0$, 4.26, 4.6 and 10.6~GeV, from Monte Carlo (MC)
simulated $e^{+} e^{-} \to \gamma_{\rm ISR} \tau^{+} \tau^{-}$,
$\tau^{-} \to \gamma_{\rm signal} \mu^{-}$, $\tau^{+} \to
anything$~\cite{kkmc1,kkmc2}. We can see that the background of $e^{+} e^{-} \to
\gamma_{\rm ISR} \tau^{+} \tau^{-}$, where the ISR photon is
misidentified as arising from $\tau$ decays, can be removed easily by
accepting as signal candidates only those photons whose energy lies
above a certain  threshold at 4.0 and
4.26~GeV without efficiency loss. Thus the expected background
level is much lower at around 4~GeV than at higher energies.

A super $\tau$-charm factory, called High Intensity Electron
Positron Accelerator (HIEPA)~\cite{HIEPA}, is being proposed in China. The
design peak luminosity is $(0.5 \sim 1)\times 10^{35}~{\rm
cm}^{-2}s^{-1}$ at $\sqrt{s}=4$~GeV with an energy range of 2 to
7~GeV. The HIEPA detector
is designed to be consists of a small-cell main drift chamber (MDC) with 48 layers,
an electro-magnetic calorimeter (EMC), a ring imaging cherenkov counter (RICH) for particle identification,
and a muon detector using muon telescope detector (MTD) method.
For the expected performance of each subdetector, see Ref.~\cite{HIEPA}.


The $\EE\to \tau^+\tau^-$ events will be
produced copiously above the $\tau\tau$ threshold. This will make
a search for the LFV process $\tau^-\to \gamma\mu^-$ possible. As
has been shown in Fig.~\ref{ISR}, the background level is expected
to be much lower at a $\tau$-charm factory than at B-factories,
and therefore it is of great interest to know what sensitivity HIEPA can
reach in searching for $\tau^-\to \gamma\mu^-$ and other LFV
processes.

In this article, we present a MC study for searching for the LFV
process $\tau^- \to \gamma \mu^-$ using a 3~fb$^{-1}$ inclusive MC
sample and estimate the sensitivity of measuring $\BR(\tau^- \to
\gamma \mu^-)$ at $\sqrt{s}=4.26$ and 4.6~GeV. We also extrapolate
our study to obtain a sensitivity on $\BR(\tau^- \to \gamma
\mu^-)$ versus integrated luminosity at 4.26~GeV. Finally we
estimate the integrated luminosity needed at HIEPA to reach the
current best upper limit of $\BR(\tau^- \to \gamma
\mu^-)<4.4\times 10^{-8}$ and to compete with the sensitivity of
$10^{-9}$ that the Belle-II experiment is expected to achieve in the near
future.

We have undertaken these studies assuming that the general performance of
the detector constructed for HIEPA will be similar to that of the BESIII
detector~\cite{BESdetector}. We do the MC simulations in
the framework of BESIII offline software system ({\sc boss})~\cite{boss}. In
the next section, we first briefly introduce the BESIII detector
and {\sc boss}, and then present the inclusive MC samples we used for
the background study for $\tau^- \to \gamma \mu^-$.

\section{BESIII Detector and Inclusive MC Samples}
\label{sec:BESMC}

The BESIII detector is a magnetic spectrometer operating at the
BEPCII Collider. The cylindrical core of the BESIII detector
consists of a helium-based main drift chamber, a plastic
scintillator time-of-flight system (TOF), and a CsI~(Tl)
electromagnetic calorimeter, which are all enclosed in a
super-conducting solenoid magnet providing a 1.0~T magnetic field.
The solenoid is supported by an octagonal flux-return yoke with
modules of resistive plate muon counters (MUC) interleaved with
steel. A detailed description of the BESIII detector is provided
in Ref.~\cite{BESdetector}.

The optimization of the event selection and the estimation of
physics backgrounds are performed through MC simulations. The {\sc
geant4}-based simulation software {\sc boost}~\cite{BOOST}
includes the geometric and material description of the BESIII
detector and the detector response and digitization models, as
well as the tracking of the detector running conditions and
performance. The analysis is performed in the framework of {\sc
boss}~\cite{boss} which takes care of the detector calibration,
event reconstruction and data storage.

The production of the charmonium resonance is simulated by the MC
event generator {\sc kkmc}~\cite{kkmc1,kkmc2}, while the decays are
generated by {\sc EvtGen}~\cite{evtgen} for known decay modes with
branching fractions being set to the PDG~\cite{PDG} world average
values, and by {\sc lundcharm}~\cite{lundcharm} for the remaining
unknown decays. The processes $e^+e^- \to \tau^+\tau^-$ and
$q\bar{q}$ ($q=u,~d,~s$) are also simulated using {\sc kkmc} based
on precise predictions of the electroweak SM. The generator {\sc
babayaga} is used to generate $e^+ e^- \to e^+ e^-$, $\mu^+
\mu^-$, $\gamma\gamma$, and $\pi^+ \pi^-$
processes~\cite{BABAYAGA}. {\sc BesTwogam} is an inclusive
generator developed from the generator {\sc Twogam} based on the
equivalent photon approximation approach and using full quantum
electrodynamics differential cross section for the process
$e^+e^-\to f \bar{f} + n\gamma$,
$f=\tau,~\mu,~d,~s,~c$~\cite{BES2GAMMA}. For more information on
the generators used at BESIII, see Ref.~\cite{evtgen}.

Signal MC samples of $e^{+} e^{-} \to \tau^{+} \tau^{-}$,
$\tau^{-} \to \gamma \mu^{-}$, $\tau^{+} \to anything$ are
generated at $\sqrt{s}=4.26$ and 4.6~GeV, to determine the
detection efficiencies.
Tables~\ref{IncMC2} and~\ref{IncMC1} summarize the generated MC
samples for background studies at $\sqrt{s}=4.26$ and 4.6~GeV, respectively. Except for
$e^{+} e^{-} \to \mu^{+} \mu^{-}$, $\tau^{+} \tau^{-}$, and $q
\bar{q}$, the sizes of the generated MC samples for the other
processes are less than 3~fb$^{-1}$ since they can be removed
completely after applying some initial event selection criteria
(discussed below) due to small production cross section or low
detection efficiency.

\begin{table}[htbp]
\caption{Generated MC samples at $\sqrt{s}=4.26$~GeV for
background study, where $\cal L$ (in fb$^{-1}$) is the
corresponding integrated luminosity.}\label{IncMC2}
\begin{tabular}{lcl}
  \hline\hline
  Process&$\cal L$&Generator \\
  \hline
  $e^{+} e^{-} \to \mu^{+} \mu^{-}$    & 3.0 &  Babayaga\\
  $e^{+} e^{-} \to \tau^{+} \tau^{-}$  & 3.0 & KKMC\\
  $e^{+} e^{-} \to q\bar{q}$ $(q = u,~d,~s)$  & 3.0 &KKMC\\
  $e^{+} e^{-} \to e^{+} e^{-}$        & 2.5 & Babayaga \\
  $e^{+} e^{-} \to \gamma \gamma$      & 2.5 & Babayaga\\
  \tabincell{c}{$e^{+} e^{-} \to \gamma_{ISR} J/\psi,\gamma_{ISR} \psi^{'}$ \\
   $\gamma_{ISR} \psi^{''},\gamma_{ISR} \psi(4040)$}
                                       & 2.5 &KKMC\\
  \tabincell{c}{$e^{+} e^{-} \to  D\bar{D}$, $D_{s}\bar{D_{s}}$,
  $D^{*}\bar{D}$ \\$D^{*}\bar{D}^{*}$, $D_{s}^{*}\bar{D}_{s}$,
  $D_{s}^{*}\bar{D}_{s}^{*}$~\footnote{For these
  $D^{(*)}\bar{D}^{(*)}$ meson pairs,
  both the charged and neutral modes are included.} }
                                       & 2.5 &KKMC\\
  $e^{+} e^{-} \to e^{+} e^{-} \gamma^{*} \gamma^{*}
  \to e^{+} e^{-}+ hadrons$            & 2.5 &BesTwogam\\
  $e^{+} e^{-} \to e^{+} e^{-} \gamma^{*} \gamma^{*}
  \to e^{+} e^{-}+ lepton~pairs$            & 2.5 & BesTwogam\\
  \hline\hline
\end{tabular}
\end{table}

\begin{table}[htbp]
\caption{Generated MC samples at $\sqrt{s}=4.6$~GeV for background
study, where $\cal L$ (in fb$^{-1}$) is the corresponding
integrated luminosity.}\label{IncMC1}
\begin{tabular}{lcl}
  \hline \hline
  Process&$\cal L$&Generator \\
  \hline
  $e^{+} e^{-} \to \mu^{+} \mu^{-}$     & 3.0 & Babayaga\\
  $e^{+} e^{-} \to \tau^{+} \tau^{-}$   & 3.0 & KKMC\\
  $e^{+} e^{-} \to q\bar{q}$ $(q = u,~d,~s)$  & 3.0 &KKMC\\
  $e^{+} e^{-} \to e^{+} e^{-}$         & 0.5 & Babayaga \\
  $e^{+} e^{-} \to \gamma \gamma$       & 0.5 & Babayaga\\
  \tabincell{c}{$ e^+ e^- \to D^{*}\bar{D}^{*}$, $D_{s}^{*}\bar{D}_{s}$,
  $D_{s}^{*}\bar{D}_{s}^{*}$\\$D\bar{D}$, $D_{s}\bar{D_{s}}$,
  $D^{*}\bar{D}$~\footnote{For these $D^{(*)}\bar{D}^{(*)}$
  meson pairs, both the charged and neutral modes are included.}}
                                        & 0.5 & KKMC\\
  $e^{+} e^{-} \to e^{+} e^{-} \gamma^{*} \gamma^{*}
  \to e^{+} e^{-} +hadrons$             & 0.5 & BesTwogam\\
  $e^{+} e^{-} \to e^{+} e^{-} \gamma^{*} \gamma^{*}
  \to e^{+} e^{-} + lepton~pairs$             & 0.5 & BesTwogam\\
  \hline\hline
\end{tabular}
\end{table}

\section{Event Selection and background analysis}
\label{sec:EveSec}

\begin{figure}[htbp]
  \includegraphics[width=0.4\textwidth]{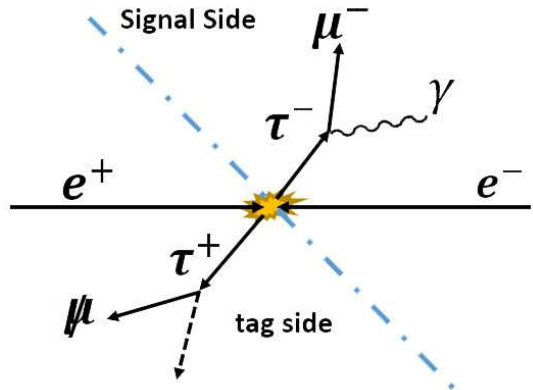}
\caption{Schematic diagram for the tagged method we used to select
$\tau^- \to \gamma \mu^-$ events. } \label{tag}
\end{figure}

We search for $\tau^- \to \gamma \mu^-$ events using a tagged
method, as depicted in Fig.~\ref{tag}, to suppress backgrounds.
The signal side is $\tau^{-} \to \gamma \mu^{-}$ while the tag
side should contain a charged particle that is not a $\mu$
(denoted as $\slashed{\mu}$), neutrino(s) and any number of
photons.

We select events that have exactly two good oppositely charged
tracks and at least one good photon.

Good charged tracks are reconstructed from MDC hits. To optimize
the momentum measurement, we select tracks in the polar angle
range $|\cos\theta|<0.93$ and require that they pass within
$\pm$10~cm of the interaction point in the beam direction and
within $\pm$1~cm of the beam line in the transverse direction.
Electromagnetic showers are reconstructed using energy deposited
in clusters of EMC crystals. The efficiency and shower energy resolution are
improved by including energy deposits in nearby TOF counters.
Showers identified as photon candidates must satisfy fiducial and
shower-quality requirements. The minimum energy is 25~MeV for
barrel showers ($|\cos\theta|<0.8$) and 50~MeV for endcap showers
($0.86<|\cos\theta|<0.92$). To exclude showers from charged
particles, a photon must be separated by at least $10^{\circ}$
from any charged track. EMC cluster timing requirements suppress
electronic noise and energy deposits unrelated to the event.

We do a kinematic fit imposing energy and momentum conservation
on the $\tau^- \to \gamma \mu^-$ for the signal side and the
$\tau^+ \to (n)\pi^0\pi^+\nu_{\tau}$ ($n=0$, 1, 2) for the tag
side hypotheses based on the total number of selected good photons
($N_{\gamma}$), i.e., $\pi^{+} \nu_{\tau}$ for  $N_{\gamma} =
1~{\rm or}~ 2$; $\pi^{0} \pi^{+} \nu_{\tau}$ for $N_{\gamma} =
3~{\rm or}~4$; and $\pi^{0} \pi^{0} \pi^{+} \nu_{\tau}$ for other
cases. Here the reconstructed momenta of two photons have been
constrained to the $\pi^0$ mass. Then we require $\chi^{2}< 13$
from the kinematic fit. We require the recoil mass of $\gamma \mu^-$
to lie within the $\tau^+$ nominal mass region ($1.70<M^{\rm
recoil}_{\gamma\mu^-}<1.81~\hbox{GeV}/c^{2}$).

For each charged track, the muon in the signal side or the
non-muon ($\slashed{\mu}$) in the tag side, the information from
the MUC and MDC is used for particle identification. Tracks with
positive penetration depth in MUC are identified as muons since
the penetrability of muon is much larger than the other charged
tracks. For the $\slashed{\mu}$ candidate, the penetration depth
in MUC is required to be less or equal zero.

\begin{figure*}[htbp]
  \includegraphics[width=0.4\textwidth]{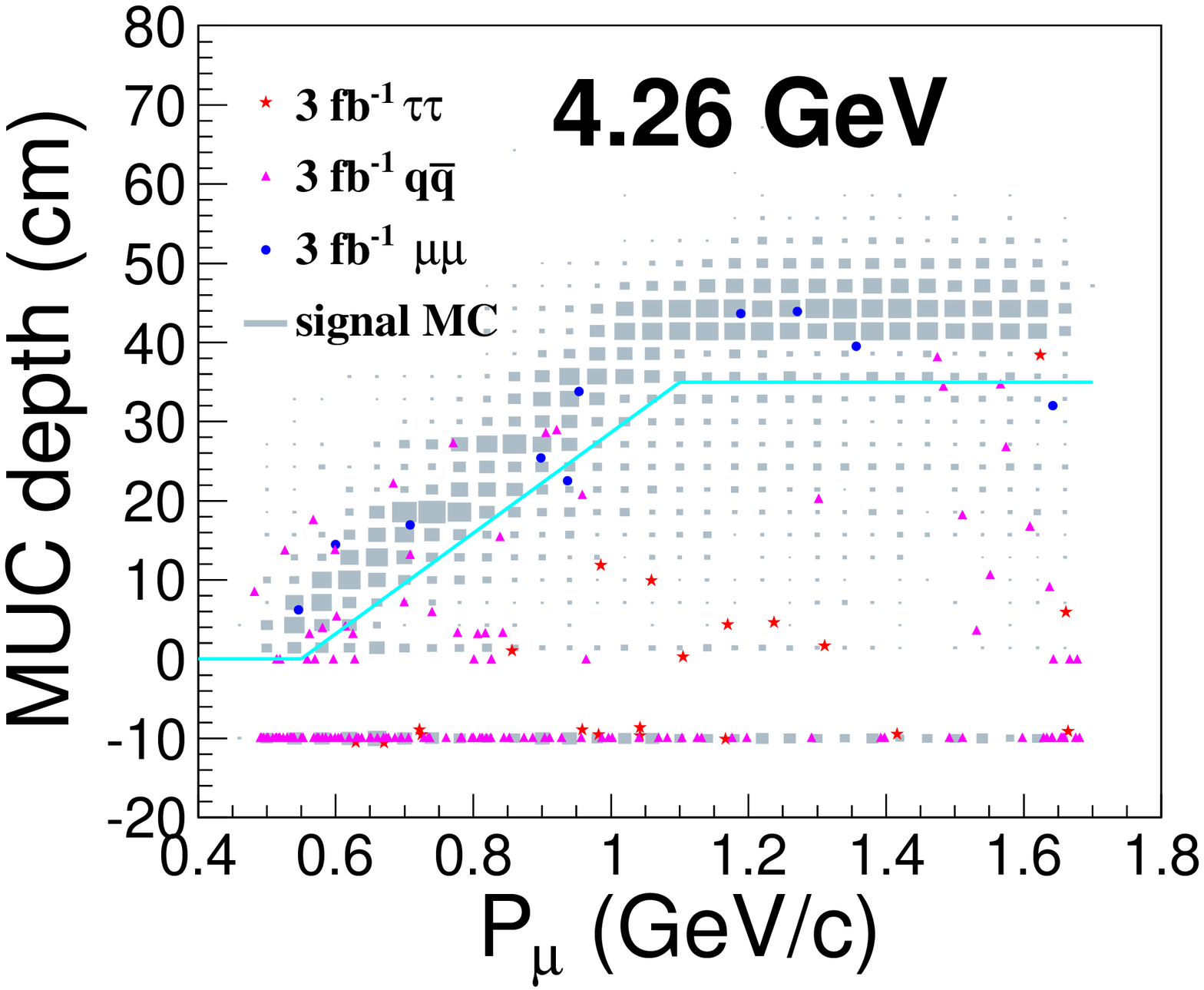}
  \includegraphics[width=0.4\textwidth]{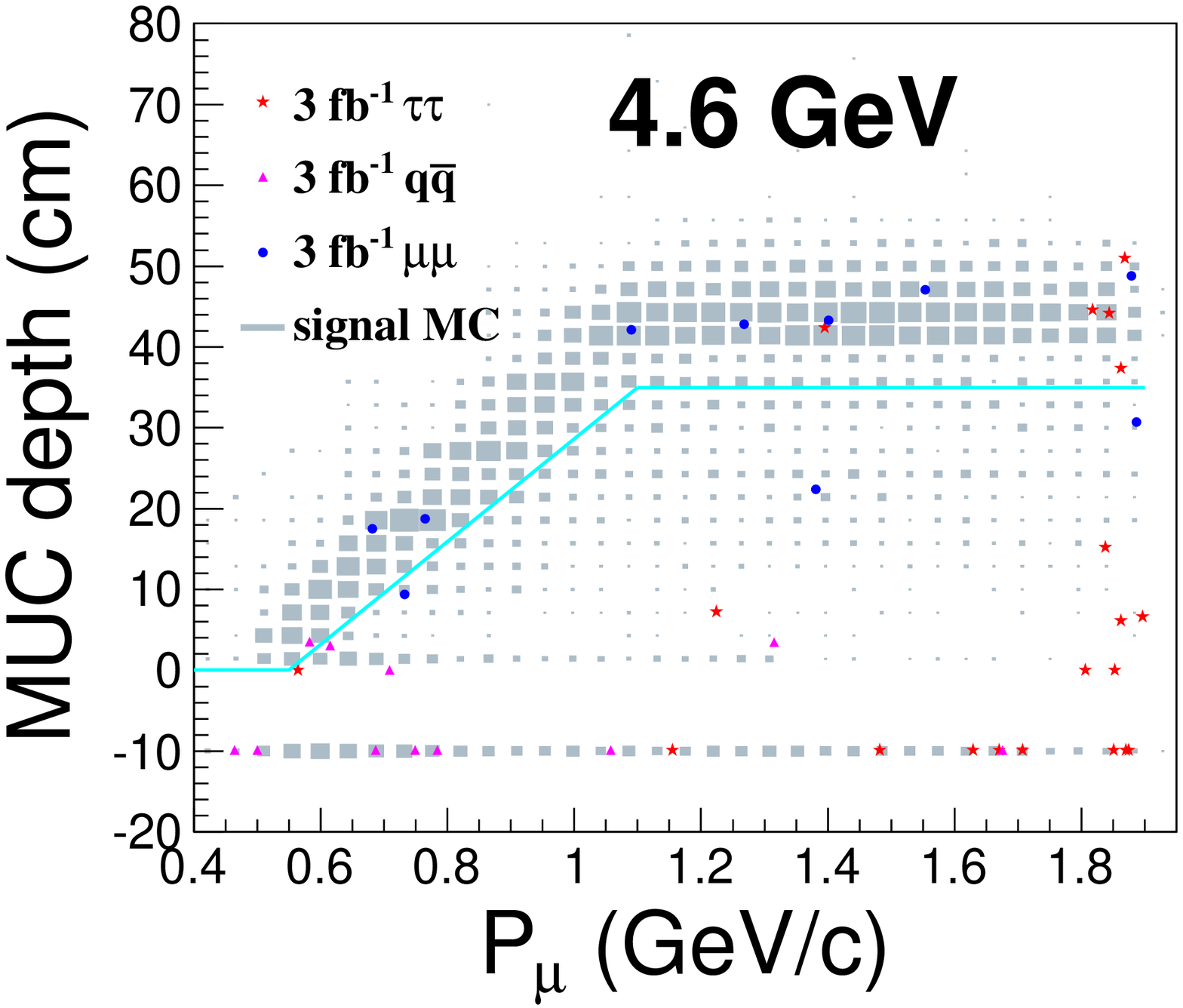}
\caption{Scatter plots of the penetration depth in MUC versus the
momentum for the muon candidates in the signal side at 4.26 (left
panel) and 4.6~GeV (right panel). The shaded boxes are from the
signal MC samples, while the marks indicate different kinds of
backgrounds indicated in the plots. The region above the solid
line is the signal region. No events survive from the other
simulated inclusive MC samples.} \label{depthvsp}
\end{figure*}

Considering the penetration depth in MUC of the muon (Dep$_{\mu}$)
varies with its momentum ($P_{\mu}$), a two-dimensional (2D)
requirement is added to further suppress the backgrounds due to
the particle misidentification (mainly from $e^{+} e^{-} \to
q\bar{q}\to \pi^{+} \pi^{-}+ X$ and $e^{+} e^{-} \to
\tau^{+} \tau^{-}$, $\tau \to \pi+ X$):
Dep$_{\mu}>(65 \times P_{\mu}-36.5)$~cm for $P_{\mu}<
1.1~\hbox{GeV}/c$ and Dep$_{\mu}>35$~cm for $P_{\mu}\geq
1.1~\hbox{GeV}$. Figure~\ref{depthvsp} shows the scatter plots of
the Dep$_{\mu}$ versus $P_{\mu}$ in the signal side at 4.26 and
4.6~GeV, where the upper region of the solid line is the required
signal region.

\begin{figure}[htbp]
  \includegraphics[width=0.23\textwidth]{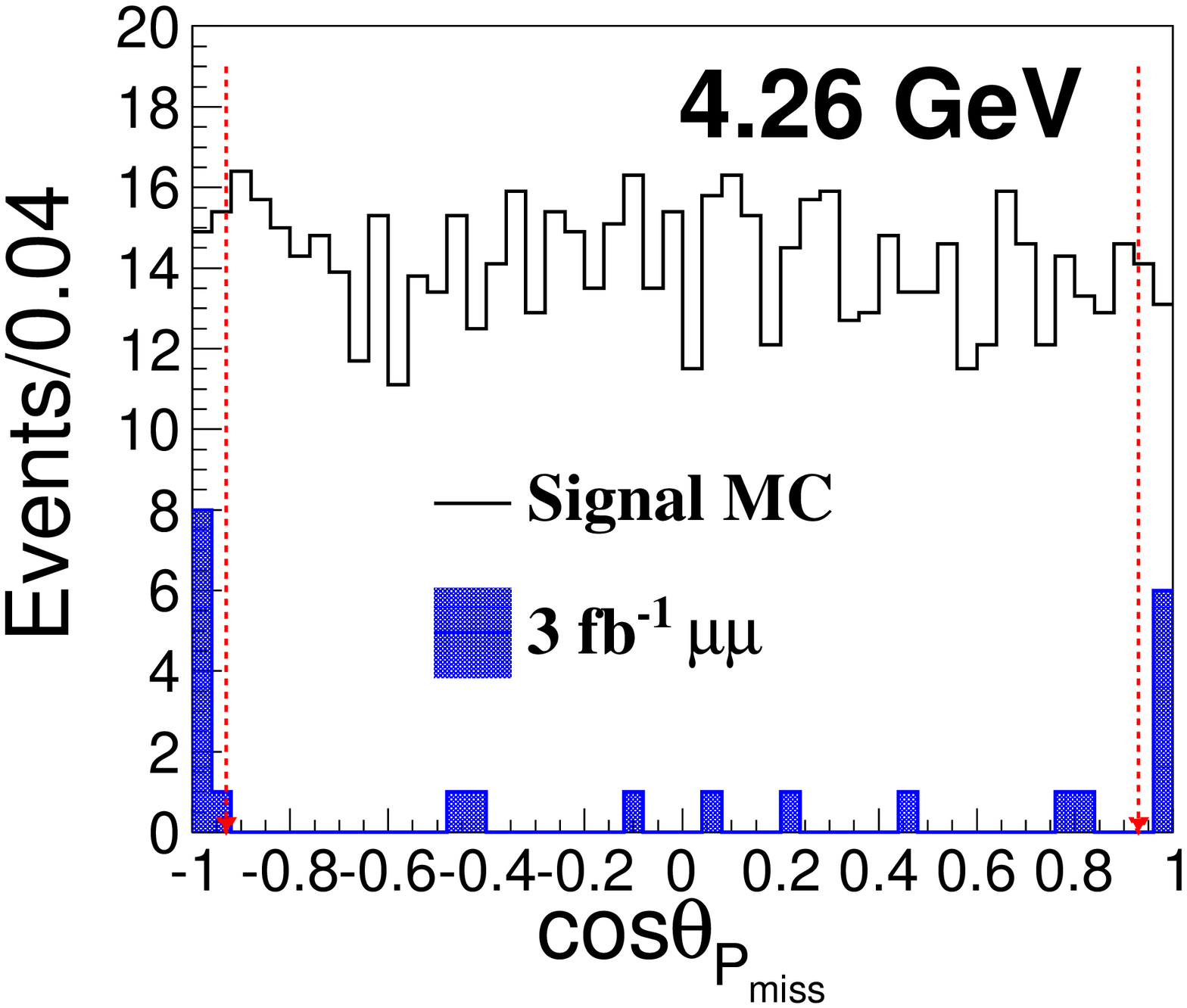}
  \includegraphics[width=0.23\textwidth]{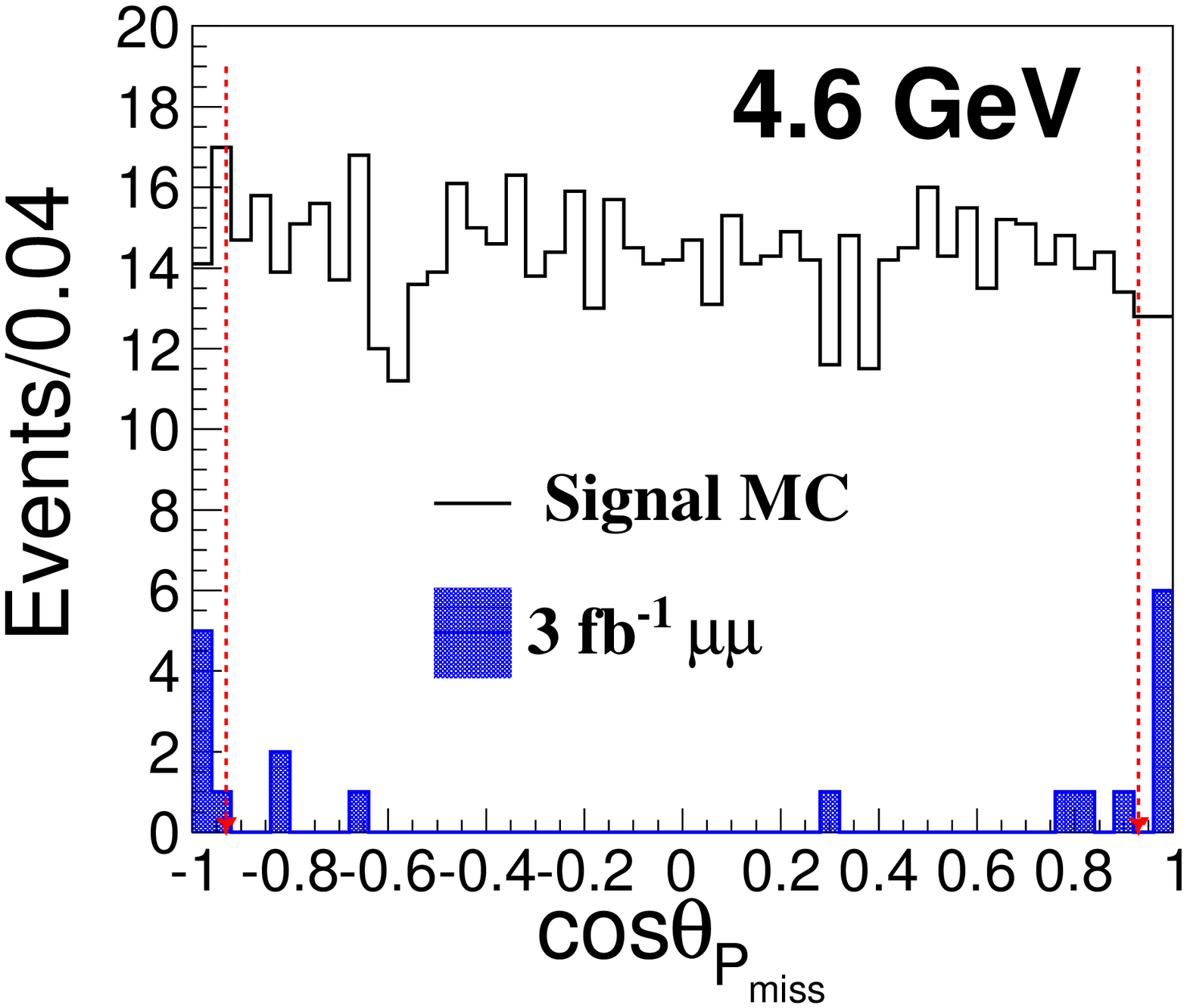}
\caption{Distributions of the $\cos\theta_{p_{\rm miss}}$ at 4.26
(left panel) and 4.6~GeV (right panel). The signal region is
between the dashed lines.} \label{COS}
\end{figure}

The backgrounds from $\mu$-pair final states are $e^{+} e^{-} \to
\gamma_{\rm ISR} \mu^{+} \mu^{-}$ with the ISR photon(s)
misidentified as the signal photon. To suppress such background
events, we require $|\cos\theta_{P_{\rm miss}}| < 0.93$, where
$P_{\rm miss}$ is calculated by subtracting the sum of momenta of
all charged tracks and the signal photon from the initial momentum of
the $e^+e^-$ system and $\theta_{P_{\rm miss}}$ is the polar angle of
$P_{\rm miss}$. Figure~\ref{COS} shows the $\cos\theta_{P_{\rm
miss}}$ distributions at 4.26 and 4.6~GeV, from which it is evident
that $e^{+} e^{-} \to \gamma_{\rm ISR} \mu^{+} \mu^{-}$ backgrounds are very
different from the signal events in polar angle distribution.

Under zero signal events assumption, we optimize the above selection
criteria to obtain the most stringent upper limits. We maximize
the figure of merit, $\cal F=\epsilon/{\rm N}_{\rm UL}$, where
$\epsilon$ is the efficiency for detecting $\tau^- \to \gamma
\mu^-$ decays obtained from the signal MC simulation and ${\rm
N}_{\rm UL}$ is the Poisson average Feldman-Cousins 90\% C.L.
upper limits on the number of expected signal events that is
obtained with the expected background and no produced
signal~\cite{UL}. The average limits ${\rm N}_{\rm UL}$ are
calculated from the number of generic MC background events
surviving the selection criteria and scaled to data size.

\begin{figure*}[htbp]
  \includegraphics[width=0.4\textwidth]{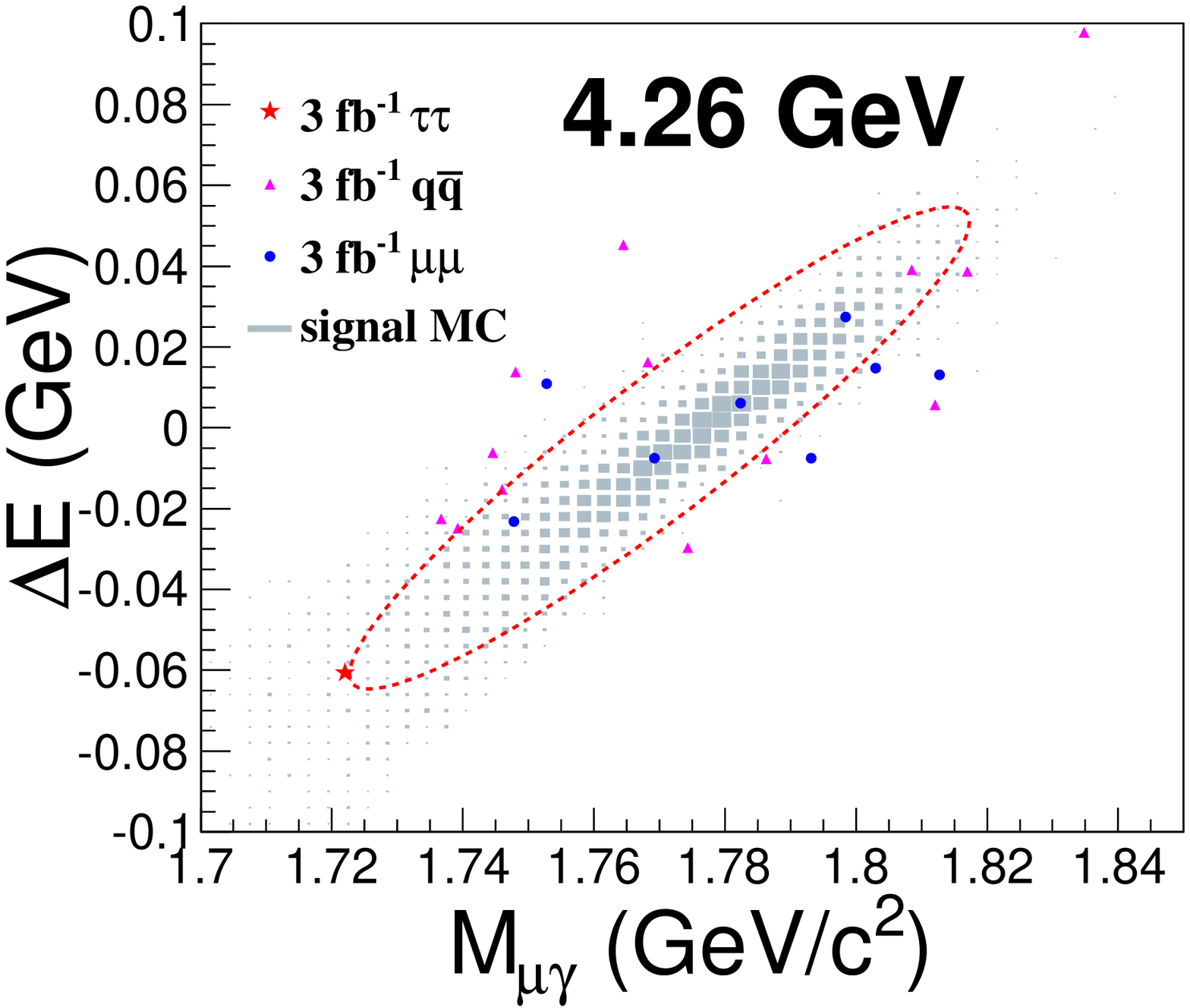}
  \includegraphics[width=0.4\textwidth]{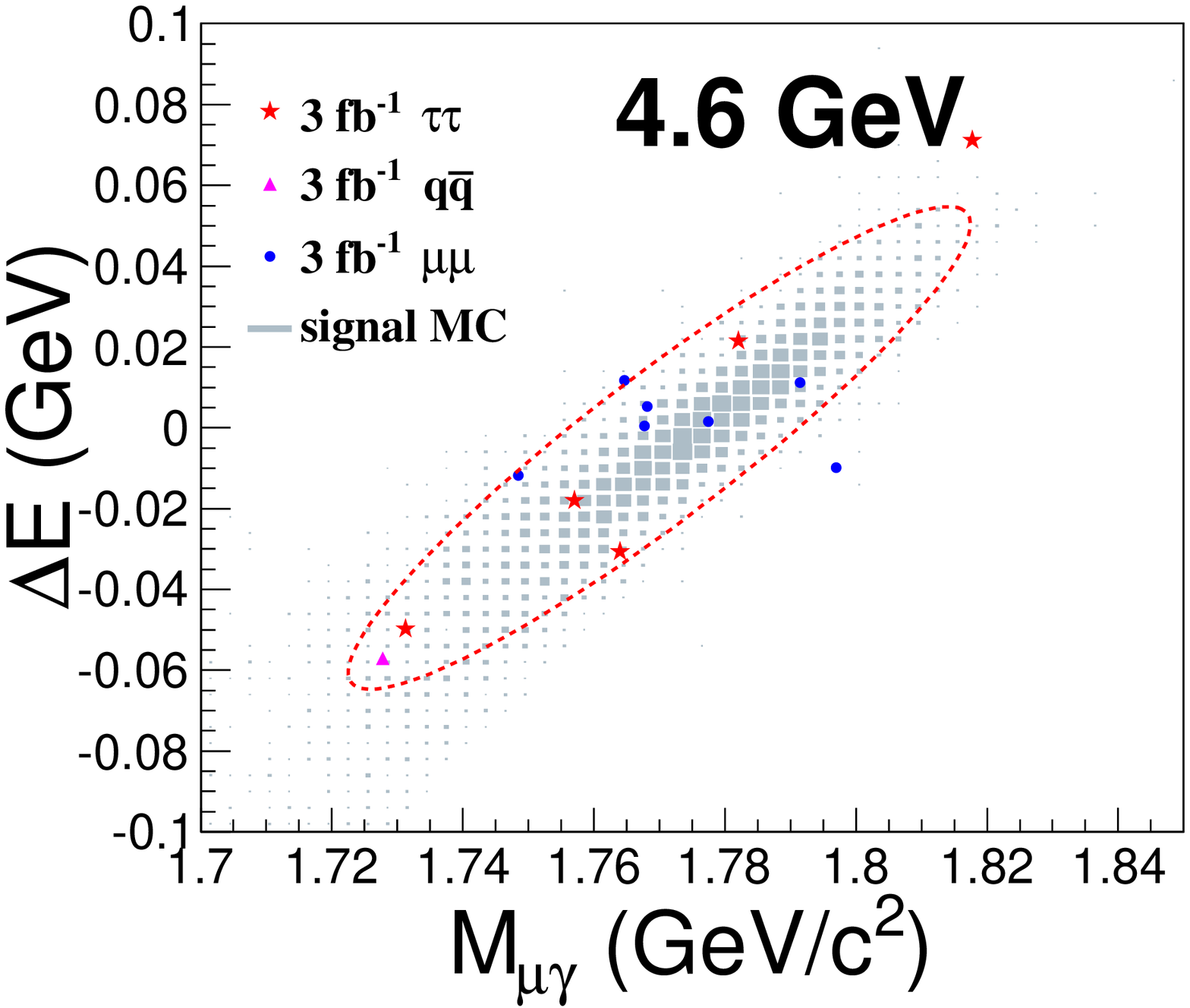}
\caption{Scatter plots of $M_{\gamma \mu^-}$ versus $\Delta E$ for
$\sqrt{s}= 4.26$ (left panel) and 4.6~GeV (right panel). The
dotted ellipses define 90\% signal regions, the shaded boxes are
from the MC simulation of signal events and the marks are from the
backgrounds shown in different colors from different sources. }
\label{Scat}
\end{figure*}

After applying all the above event selection, signal candidates
are examined in the 2D space of the invariant mass ($M_{\gamma
\mu^-}$) and the difference of their energy from the beam energy
in the $e^+e^-$ CM system ($\Delta E$). For a signal, $M_{\gamma
\mu^-}$ will center on the the $\tau$-lepton mass and $\Delta E$
on zero.  We define a  elliptical signal region which contains
90\% of signal events according to signal MC simulations.
Figure~\ref{Scat} shows the scatter plots of $M_{\gamma \mu^-}$
versus $\Delta E$ at 4.26 and 4.6~GeV, where the shaded boxes are
from the MC signal simulations and the dots are from the
backgrounds shown in different colors from different sources.
With the generated integrated
luminosity of 3 fb$^{-1}$, 5 and 11 background events survive
in the signal ellipses at 4.26 and 4.6~GeV, respectively.

\section{Results and Discussion}
\label{sec:summary}

We determine upper limit on the branching fraction $\BR(\tau^-\to
\gamma \mu^-)$ at 90\% C.L. with the following formula:
\begin{equation*}
\BR(\tau^{-} \to \gamma \mu^{-}) <
\frac{N_{\rm UL}\times |1-\Pi(s)|^2}{2\epsilon \times
\sigma(e^+e^-\to \tau^+ \tau^-)\times {\cal L} \times (1+\delta)},
\label{FUN}
\end{equation*}
where $N_{\rm UL}$, $\epsilon$, $1+\delta$, $|1-\Pi(s)|^2$, and
$\cal L$ are the Poisson average Feldman-Cousins 90\% C.L. upper
limit on the number of expected signal events mentioned
above~\cite{UL}, the detection efficiency, the radiative
correction factor obtained from the ratio of the
$e^+e^- \to \tau^+\tau^-$ cross sections with the
ISR turned on and off in {\sc kkmc}~\cite{ISRF}
generator, the vacuum polarization factor, and the
integrated luminosity, respectively.

The upper limits on the expected signal events are
7.50 and 10.6; the Born cross
sections $\sigma(e^{+} e^{-} \to \tau^{+} \tau^{-})$ are 3.56 and
3.38~nb; the radiative correction factors $1+\delta$ are 0.96 and
0.98, the detection efficiencies are 5.92\% and 5.90\%, and the
vacuum polarization factors $|1-\Pi(s)|^2$ are both 0.98~\cite{vp1,vp2,vp3,vp4}
for $\sqrt{s}=4.26$ and 4.6~GeV, respectively.

With the integrated luminosity of
3~fb$^{-1}$ at 4.26 and 4.6~GeV, the upper limits on $\BR(\tau^-
\to \gamma \mu^-)$ are determined to be less than $6.1
\times 10^{-6}$, and $8.9 \times 10^{-6}$ for $\sqrt{s}=4.26$ and
4.6~GeV, respectively.

In calculating the above upper limits, we count the number of
expected background events in the 90\% signal region. If we take the
68.3\% signal region (1$\sigma$), 4 and 6 background events survive,
and the detection efficiencies become 4.6\% and 4.55\%, respectively,
at  $\sqrt{s}=4.26$ and 4.6~GeV. The upper limits on the expected signal events are
6.6 and 8.3, which correspond to $6.9\times 10^{-6}$ and $9.0\times 10^{-6}$ upper limits on $\BR(\tau^-
\to \gamma \mu^-)$ at 4.26 GeV and 4.6 GeV, respectively.
The signal region selection could be further optimized with much larger
MC inclusive samples in the future.

\begin{figure}[htbp]
  \includegraphics[width=0.4\textwidth]{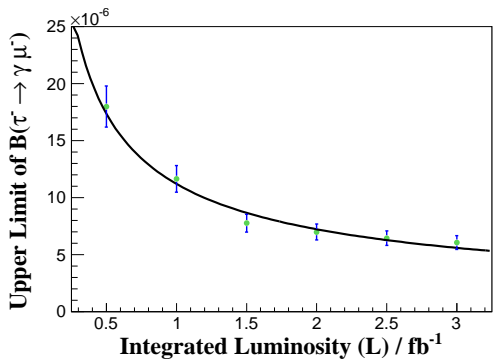}
\caption{The estimated upper limit on $\BR(\tau^- \to \gamma
\mu^-)$ at 90\% C.L. versus the integrated luminosity at
$\sqrt{s}=4.26$~GeV. The solid line shows the fitted result with a
function of $\alpha \cal{L}^\beta$. } \label{FIT}
\end{figure}

To estimate how large a data sample is needed for HIEPA to achieve
the current best upper limit on $\BR(\tau^- \to \gamma \mu^-)$, we
calculate $\BR(\tau^{-} \to \gamma \mu^{-})$ under the
assumptions of the integrated luminosity of 0.5, 1, 1.5, 2 and 2.5
fb$^{-1}$ using the same method. Since the sensitivity of
$\BR(\tau^- \to \gamma \mu^-)$ at 4.26 GeV is better than that
at 4.6 GeV with the same integrated luminosity, here we just
study the sensitivity versus the integrated luminosity at 4.26 GeV.
Figure~\ref{FIT} shows the
estimated 90\% C.L. upper limits versus the integrated luminosity
at $\sqrt{s}=4.26$ GeV.
The solid line shows the fitted result with a function of $\alpha
\cal{L}^\beta$, where $\alpha$ and $\beta$ are free parameters.
From the fit, we obtain $\beta=-0.632\pm0.072$.
With the fitted results, HIEPA needs to take at least a
6.34~ab$^{-1}$ data sample to
obtain the current best upper limit of $4.4\times 10^{-8}$.

The Belle-II experiment is going to take data beginning in 2018 with a
design luminosity of $8\times 10^{35}~{\rm cm}^{-2}~{\rm
s}^{-1}$ at the $\Upsilon(4S)$ peak, and its integrated luminosity
will reach 50~ab$^{-1}$ by 2024. With this sample $5\times10^{10}$
$\tau$-pair events will be accumulated, and a sensitivity of $1\times 10^{-9}$
is expected for $\BR(\tau^- \to \gamma \mu^-)$ if the
Belle-II signal-to-background conditions are the same as that
of Belle~\cite{BELLE}. To achieve a similar sensitivity, HIEPA needs to
take at least $2051$~ab$^{-1}$
data sample. It means HIEPA can not compete with Belle-II without
improving the detector performance over BESIII detector.

The remaining backgrounds are due to the $\mu$ and $\pi$
misidentification in the framework of BESIII offline software
system. Fortunately, the expected $\mu/\pi$ separation power will
increase a lot ($>10$ times) at HIEPA compared to
BESIII~\cite{HIEPA}. Therefore, the particle misidentification
backgrounds can be further suppressed at HIEPA significantly.
Assuming negligible background level the 90\% C.L. upper limit
on $\BR(\tau^- \to \gamma \mu^-)$ is expected to be proportional
to $1/\cal {L}$, to reach $1\times 10^{-9}$ sensitivity HIEPA
needs to take at least a $18.3$~ab$^{-1}$ data sample for a design peak luminosity of
$0.5\times 10^{35}~{\rm cm}^{-2}~{\rm s}^{-1}$ at 4.26~GeV.

\section*{ACKNOWLEDGMENTS}

Y. B. Li thanks Profs. Jing-Zhi Zhang, Rong-Gang Ping and Mr.
Lian-Jin Wu for their helpful discussions and cross checks on MUC
information and generator, also thanks Mr. Hong-Rong Qi for his
help on the optimization of event selection. This work is
supported in part by the National Natural Science Foundation of
China Under Contract No. 11235011, National Key Basic Research
Program of China under Contact No. 2015CB856701, and CAS center
for Excellence in Particle Physics (CCEPP).


\begin{thebibliography}{0}
\expandafter\ifx\csname natexlab\endcsname\relax\def\natexlab#1{#1}\fi
\expandafter\ifx\csname bibnamefont\endcsname\relax
  \def\bibnamefont#1{#1}\fi
\expandafter\ifx\csname bibfnamefont\endcsname\relax
  \def\bibfnamefont#1{#1}\fi
\expandafter\ifx\csname citenamefont\endcsname\relax
  \def\citenamefont#1{#1}\fi
\expandafter\ifx\csname url\endcsname\relax
  \def\url#1{\texttt{#1}}\fi
\expandafter\ifx\csname urlprefix\endcsname\relax\def\urlprefix{URL }\fi
\providecommand{\bibinfo}[2]{#2}
\providecommand{\eprint}[2][]{\url{#2}}

\end{thebibliography}


\begin{thebibliography}{99}

\bibitem{SM}  B. W. Lee and R. E. Shrock, Natural Suppression of Symmetry Violation in Gauge Theories: Muon - Lepton and Electron Lepton Number Nonconservation, Phys. Rev. D
{\bf 16}, 1444 (1977).

\bibitem{PDG} K. A. Olive, K. Agashe, C. Amsler {\it et al.} (Particle Data Group), Review of Particle Physics , Chin. Phys. C
{\bf 38}, 090001 (2014) and 2015 update.

\bibitem{SSM} A. Brignole and A. Rossi, Anatomy and phenomenology of mu-tau lepton flavor violation in the MSSM, Nucl. Phys. B
{\bf 701}, 3 (2004).

\bibitem{grand} L. Calibbi, A. Faccia, A. Masiero and S.K. Vempati, Lepton flavour violation from SUSY-GUTs: Where do we stand for MEG, PRISM/PRIME and a super flavour factory,
 Phys. Rev. D {\bf 74}, 116002 (2006).

\bibitem{SEESAW} J. R. Eillis, J. Hisano, M. Raidl and Y. Shimizu, A New parametrization of the seesaw mechanism and applications in supersymmetric models,
Phys, Rev. D {\bf 66}, 115013 (2002).

\bibitem{review} R. H. Bernstein and P. S. Cooper, Charged Lepton Flavor Violation: An Experimenter's Guide, Phys. Rept. {\bf 532}, 27 (2013).

\bibitem{footnote} Charge conjugate mode is included throughout
this paper, the sensitivity in this paper is for combined
$\tau^-\to \gamma\mu^-$ and $\tau^+\to \gamma\mu^+$.

\bibitem{mue1} W. Bertl, R. Engfer, E. A. Hermes, {\it et al.} (SINDRUM II Collaboration), A Search for muon to electron conversion in muonic gold, Eur. Phys. J. C {\bf 47} 337-346 (2006).

\bibitem{mue2} R. K. Kutschke, The Mu2e Experiment at Fermilab, arXiv:1112.0242.

\bibitem{babar} B. Aubert, Y. Karyotakis, J.P. Lees, {\it et al.} (BaBar Collaboration), Searches for Lepton Flavor Violation in the Decays
$\tau^{\pm} \to e^{\pm} \gamma$ and $\tau^{\pm} \to e^{\pm} \gamma$,
Phys. Rev. Lett. {\bf 104}, 021802 (2010).


\bibitem{kkmc1} S. Jadach, B. F. L. Ward and Z. Was, The Precision Monte Carlo event generator K K for two fermion final states in e+ e- collisions, Comp. Phys.
Commu. {\bf 130}, 260 (2000)
\bibitem{kkmc2} S. Jadach, B. F. L. Ward and Z. Was, Coherent exclusive exponentiation for precision Monte Carlo calculations, Phys. Rev. D {\bf 63}, 113009 (2001).



\bibitem{HIEPA} C. Li's talk ``Overview of the Detector at HIEPAF" given
at ``Workshop on Physics at Future High Intensity Collider at 2-7GeV in China, 13 - 16 January, 2015, Hefei, China".

\bibitem{BESdetector} M. Ablikim, Z. H. An, J. Z. Bai, {\it et al.} (BESIII Collaboration), Design and Construction of the BESIII Detector,
Nucl. Instrum. Meth. A {\bf 614}, 345 (2010).


\bibitem{boss} W. D. Li, H. Liu, Z. Deng, {\it et al.}, The offline software for the BESIII experiment, Proc. CHEP {\bf 27} 225 (2006)

\bibitem{BOOST} Z. Y. Deng, G. F. Cao, C. D. Fu, {\it et al.}, Object-Oriented BESIII Detector Simulation System, HEP $\&$ NP {\bf 30(05)}, 371-377 (2006).

\bibitem{evtgen} R. G. Ping, Event generators at BESIII, Chin. Phys. C
{\bf 32}, 599 (2008).

\bibitem{lundcharm} J. C. Chen, G. S. Huang, X. R. Qi,
D. H. Zhang and Y. S. Zhu, Event generator for $J/\psi$ and $\psi(2S)$ decay,  Phys. Rev. D {\bf62}, 034003 (2000).

\bibitem{BABAYAGA} G. Balossini, C. M. C. Calame and G. Montagna, Matching perturbative and parton shower corrections to Bhabha process at flavour factories,
Nucl. Phys. B {\bf758}, 227 (2006).

\bibitem{BES2GAMMA} S. Nova, A. Olchevski and T. Todorov, A Monte Carlo event generator for two photon physics,
DELPHI Note 90-35 152 (1990).

\bibitem{UL} G. J. Feldman and R. D. Cousin, A Unified approach to the classical statistical analysis of small signals,
Phys. Rev. D {\bf 57}, 3873 (1998).
\bibitem{ISRF} E. A. Kuraev and V. S. Fadin, On Radiative Corrections to $\EE$ Single Photon Annihilation at High-Energy, Yad. Fiz. {\bf41}, 733 (1985).
\bibitem{vp1}  F.~Jegerlehner, Hadronic Contributions to Electroweak Parameter Shifts: A Detailed Analysis, Z. Phys. C {\bf 32}, 195 (1986).
\bibitem{vp2} S.~Eidelman and F.~Jegerlehner, Hadronic contributions to g-2 of the leptons and to the effective fine structure constant alpha (M(z)$^{2}$), Z. Phys. C {\bf 67}, 585 (1995).
\bibitem{vp3} F.~Jegerlehner, The Role of sigma($\EE \to hadrons$) in precision tests of the standard model, Nucl. Phys. Proc. Suppl. {\bf 131}, 213 (2004).
\bibitem{vp4} F.~Jegerlehner, Precision measurements of sigma(hadronic) for $\alpha$(eff)(E) at ILC energies and (g-2)($\mu$), Nucl. Phys. Proc. Suppl. {\bf 162}, 22 (2006).

\bibitem{BELLE} T. Aushev, W. Bartel, A. Bondar, {\it et al.}, Physics at Super-B Factory, arXiv:1002.5012.



\end{thebibliography}
\end{document}